\documentclass[12pt,epsf]{article}
\usepackage{epsfig}
\usepackage{amssymb}  
\usepackage{amsthm}
\usepackage{amsmath}
\let\eqref\cref
\newcommand{\onefigure}[2]{\begin{figure}[htbp]
\begin{center}\leavevmode\epsfbox{#1.eps}\end{center}\caption{#2\label{#1}}
\end{figure}}

\renewcommand{\thanks}[1]{\footnote{#1}} 

\renewcommand{\theequation}{\thesection.\arabic{equation}}
\newcommand{\be}{\begin{equation}}
\newcommand{\ee}{\end{equation}}
\newcommand{\bea}{\begin{eqnarray}}
\newcommand{\eea}{\end{eqnarray}}
\newcommand{\half}{{1\over 2}}

\def\stacksymbols#1#2#3#4{\def\theguybelow{#2}
\def\verticalposition{\lower#3pt}
\def\spacingwithinsymbol{\baselineskip0pt\lineskip#4pt}
\mathrel{\mathpalette\intermediary#1}}
\def\intermediary#1#2{\verticalposition\vbox{\spacingwithinsymbol
\everycr={}\tabskip0pt
\halign{$\mathsurround0pt#1\hfil##\hfil$\crcr#2\crcr
\theguybelow\crcr}}}

\newcommand{\goesto}[2]{\stacksymbols{\Longrightarrow}{{}_{#1 \rightarrow #2}}{4}{0.1} }
\newcommand{\goeslike}[3]{\stacksymbols{\Longrightarrow}{#1 #2 #3}{4}{0.1} }

\begin{document}

\pagestyle{empty}

\bigskip\bigskip
\begin{center}
{\bf \large Non-Perturbative Unitary Mixing of Bosonic Causal Spinors}
\end{center}

\begin{center}
James Lindesay\footnote{e-mail address, jllindesay@howard.edu} \\
Computational Physics Laboratory \\
Howard University,
Washington, D.C. 20059 
\end{center}
\bigskip

\begin{abstract} 
The fundamental fermion representations of causal spinor fields
have previously been demonstrated to describe free Dirac fermions, as well as incorporate
\emph{only} the observed degrees of freedom for local gauge invariance.  In this paper, the first non-trivial
boson representations of causal spinors will be presented.  The general unitary mixing of degenerate bosonic
spinors will be developed, and then applied to electro-weak bosons satisfying appropriate
kinematic constraints.  The resulting mass ratios are seen to be consistent with
reported measured values for these bosons.

\end{abstract}

\bigskip 

\setcounter{equation}{0}
\section{Introduction}
\indent \indent
Fundamental formulations of quantum dynamics must incorporate conservation of
momenta, positive-definite energies, and angular momenta, as well as the maintenance
of various internal quantum numbers through unitarity and pairwise
creations/annihilations, 
when appropriate.  Furthermore, correspondence of quantum interactions with classical
formulations requires cluster decomposability, as well as being able to establish
causal relationships between interactions.

Causality allows for space-like correlations in quantum systems
(including `spooky' entanglements, etc.) without
allowing space-like (faster-than-light) communications for separations
 satisfying $(x^\mu - y^\mu) \eta_{\mu \nu} (x^\nu - y^\nu)>0$,
where two events have coordinates $\vec{x}$ and $\vec{y}$.  At the microscopic
level, this requires that fermion/boson
fields must anti-commute/commute outside of the light-cone, i.e.,
$[\hat{\mathbf{\psi}}_{\gamma}^{(\Gamma)}(\vec{x}),
\hat{\mathbf{\psi}}_{\gamma}^{(\Gamma)}(\vec{y})]_{\pm}=0$ for $\vec{x}-\vec{y}$
space-like. This requirement connects the quantum statistics to the spin, such
that $(\pm)=-(-1)^{2 J}=-(-1)^{2 \Gamma}$.

At their most primary level, causal spinor fields\cite{JLLSF13} necessarily involve two particle types
that pair to satisfy microscopic causality, while transforming under the usual translations, rotations,
and Lorentz boosts of the Poincare' group.  The fundamental representation spinors
are 4-dimensional, satisfying the usual Dirac algebra\cite{Dirac}.  In addition to the generators
of group transformations, there remain exactly 12 additional hermitian generators of internal
(non space-time) symmetries that include a U(1), an SU(2), and for massive fermions
an SU(3) that has been shown to \emph{require} mixing 3 generations of mass eigenstates\cite{JLSU3SU2U1}.
The interactions generated by local gauge bosons associated with these generators must be
incorporated into any geometric interaction via covariant curvilinear coordinate
transformations from the flat space-time translations of the group.  Furthermore,
the non-vanishing structure constants of the group directly \emph{constructs} the Minkowski metric
\emph{within} the group parameter space defining group invariants.

The spinor field equation requires the form $\hat{\Gamma}^\mu \: \hat{P}_\mu$ to
be a Lorentz scalar operation, satisfying
\be
\mathbf{\Gamma}^\beta \cdot {\hbar \over i} { \partial \over \partial x^\beta} \,
\hat{\mathbf{\psi}}_{\gamma}^{(\Gamma)}
(\vec{x}) = -\gamma m  c  \: \hat{\mathbf{\psi}}_{\gamma}^{(\Gamma)}(\vec{x}) ,
\label{CausalSpinorFieldEqn}
\ee
where $m>0$, and $\mathbf{\Gamma}^\beta$ are the
finite dimensional matrix representations of the operators $\hat{\Gamma}^\beta$. 
In (\ref{CausalSpinorFieldEqn}), the particle mass squared $m^2$ are eigenvalues of the Casimir
operator which labels the unitary representations of the (extended) Poincare group algebra. 
For massive particles, the label $\gamma$ is the particular eigenvalue of the hermitian matrix $\mathbf{\Gamma}^0$, and for the $\Gamma=\half$ (fermion) representation, the Dirac matrices
are twice $\mathbf{\Gamma}^\beta$.  In this paper, the $\Gamma=1$ boson representation will be
explored.  In particular, since this 10-dimensioonal representation inherently combines scalars with vectors,
special attention will be given to an examination of electro-weak gauge bosons. 
In what follows, natural units with $\hbar=1, c=1$ will be utilized.


\section{Bosonic Spinors}
\subsection{General form of $\Gamma=1$ causal spinors}
\indent \indent
The ten-dimensional representations of the $\mathbf{\Gamma}^\beta$ matrices are presented in the appendix
(\ref{GammaMatrices}).  In what follows, motions will be confined to the z-axis,
since eigenspinors with general momenta can be obtained through a simple rotation. 
The eigenspinors satisfy
\be
\mathbf{\Gamma}^\beta p_\beta \,
 \mathbf{\Psi}_{\gamma}^{(1)} ( \vec{p},J, s_z)
 = -\gamma m \:  \mathbf{\Psi}_{\gamma}^{(1)} ( \vec{p},J, s_z),
\label{CausalEigenSpinorEqn}
\ee
where $\gamma$  and $J$ are integers no greater than $\Gamma=1$.
Microscopic causality requires\cite{JLFQG}\cite{WeinbergQTF} that the space-time dependent boson fields combine
a spinor labeled by $\gamma$ with one labeled by $-\gamma$,
(i.e., an appropriate combining of particle with anti-particle).
Relevant hermitian normalized eigenspinors for z-moving systems are presented
in Equations (\ref{DegenerateSpinors}) and (\ref{TypicalSpinors}) in the appendix. 

The ten standard state ($\vec{p}\rightarrow\{m,0,0,0  \}$ at rest) spinors include a scalar with $J=0,\gamma=0$,
as well as vector triplets with $J=1, s_z=+1,0,-1$ for each possible value of
$\gamma=+1,0,-1$.  It is worth mentioning that this also corresponds with the number of charged and
neutral weak bosons, including the Higgs.  In the next section, the unitary mixing of
the spinors will be examined.


\subsection{Unitary mixing of spinors}
\indent \indent
A noteworthy property of (\ref{CausalSpinorFieldEqn}) 
that is \emph{not} true of the Dirac representation
is that fields with $\gamma=0$ are
all degenerate regardless of mass, including massless fields.  This implies that
unitary mixing of degenerate fields can result in new fields with differing masses
that continue to satisfy the field equation.
Furthermore, for massless particles $m \rightarrow 0$,
fields given by $\hat{A}_{\gamma}^{(\Gamma) \, \mu}(\vec{x}) \equiv
\mathbf{\Gamma}^\mu \hat{\mathbf{\psi}}_{\gamma}^{(\Gamma)}(\vec{x})$
define contravariant field components that inherently satisfy the Lorentz gauge condition
${\partial \over \partial x^\mu}\hat{A}_{\gamma}^{(\Gamma) \, \mu}(\vec{x})=0$.

It is thus of interest to examine the mixing of $\Gamma=1$ vector (J=1) and
scalar (J=0) bosons, especially in relation to the mixing of electro-weak bosons.
No mixing that preserves normalization between spinors of differing masses
with $\gamma \neq 0$ can be found.  However, orthogonal mixing
of the degenerate $\gamma=0$ spinors in (\ref{DegenerateSpinors}) \emph{can}
be developed.  Using the identifications
$m_B\rightarrow i Q_B$, with
\be
p_B \rightarrow \mp {Q_B \over m_W} \sqrt{m_W^2 + p_W^2} \quad , \quad
\sqrt{m_B^2 + p_B^2} \rightarrow \pm {Q_B \over m_W} p_W,
\ee
a time-like $W_0$ and orthogonal space-like $B_0$ are described. 
In this expression, the $\pm$ sign corresponds with the sign of $p_W$.
Thus, the four $\gamma=0$ degenerate $W_0$ and $B_0$ spinors
can mix to generate new spinor forms $Z$ and $A$ according to
\be
\begin{array}{r}
\cos \phi_{WB} \: \mathbf{\Psi}_{0}^{(1)} ( \vec{p}_W,J, s_z) +
\sin \phi_{WB} \: \mathbf{\Psi}_{0}^{(1)} ( \vec{p}_B,J, s_z) =
\mathbf{\Psi}_{0}^{(1)} ( \vec{p}_Z,J, s_z), \\
-\sin \phi_{WB} \: \mathbf{\Psi}_{0}^{(1)} ( \vec{p}_W,J, s_z) +
\cos \phi_{WB} \: \mathbf{\Psi}_{0}^{(1)} ( \vec{p}_B,J, s_z) =
\mathbf{\Psi}_{0}^{(1)} ( \vec{p}_A,J, s_z),
\end{array}
\label{WBmixingEqn}
\ee
where for z-moving systems, the parameter $s_z$ labels the helicity of the particle.

In this development, it is sufficient to examine only z-moving momenta $p$,
since general motions can be defined by a simple rotation into the direction of motion. 
It is convenient to define the dimensionless parameter $\zeta_m$ as follows:
\be
\zeta_m \equiv \sin^{-1}\left (
p_m \over \sqrt{m^2 + 2 p_m^2}
\right ), \qquad p_m = m {\sin \zeta_m \over \sqrt{\cos (2 \zeta_m)}}.
\label{zetamEqn}
\ee
Using this definition, a typical $\gamma=0$ normalized spinor demonstrated
in appendix Eq. (\ref{DegenerateSpinors}) takes the form
\be
\mathbf{\Psi}_{0}^{(1)} ( \vec{p}_m,1, 1)=\left (
\begin{array}{c}
0 \\ \sin \zeta_m \over \sqrt{2}  \\ 0 \\ 0 \\
\cos \zeta_m \\ 0 \\ 0 \\ \sin \zeta_m \over \sqrt{2} \\ 0 \\ 0
\end{array} \right ).
\ee
For massive particles, $-{\pi \over 4}<\zeta_m<{\pi \over 4}$, while massless particles
have $\zeta_0= \pm {\pi \over 4}$, and space-like 4-momenta satisfy
$ {\pi \over 4}<|\zeta_{i Q}|< {\pi \over 2}$.
Expressed using this parameterization, the mixing satisfies
\be
\sin \phi_{WB}=\sin (\zeta W - \zeta_Z),
\label{PhiWBeqn}
\ee
where $\zeta W$ is defined using the superposition of the $\gamma=0$
spinors in (\ref{WBmixingEqn}). 
There are alternative expressions of this mixing angle in terms of 
$\zeta B$ and $\zeta_A$ (with $|\zeta W-\zeta B|={\pi \over 2}$ for orthogonality), 
but the above form is most convenient
for the kinematic identifications to be discussed in what follows.

Under (active) Lorentz boosts $\gamma_{LT}\equiv {1 \over \sqrt{1-\beta_{LT}^2}}$,
the parameters $\zeta_m$ transform according to
\be
\zeta_m \goeslike{}{\beta_{LT}}{} \tilde{\zeta}_m =
\sin^{-1}  \left (
{\sin \zeta_m + \beta_{LT} \cos \zeta_m \over
\sqrt{1+2 \beta_{LT} \sin (2 \zeta_m) +\beta_{LT}^2 }}
\right ).
\label{LTofZetaEqn}
\ee
Since this expression is completely independent of mass $m$, this means that
any two different masses $m_A \neq m_B$ with
equal  $\zeta$ values $\zeta_A =\zeta_B$ are necessarily \emph{co-moving},
but displaying differing momenta.  In particular, the Lorentz transformation that boosts a mass
with finite $\zeta_m$ to rest is given by $\beta_{LT}^{rest}=-\tan \zeta_m$.

General $W_0 - B_0$ spinor mixing kinematics can be expressed in terms of Lorentz invariants by defining
$(P_W^\mu + P_B^\mu) \, \eta_{\mu \nu} \,  (P_W^\mu + P_B^\mu) \equiv -M^2$ using
\be
\zeta W ^M _{Q_B}=\sin^{-1} \left (
\sqrt{  -2 m_W Q_B + \sqrt{M^4+2 M^2 (Q_B^2-m_W^2 ) +(Q_B^2 +m_W^2)^2   }   } \over
\sqrt{2} \left ( (M^2-m_W^2)^2 + 2 (M^2 +m_W^2) Q_B^2 + Q_B^4    \right )^{1 \over 4}
\right )
\label{zetaWBmixEqn}
\ee
This defines the spinor mixing in (\ref{WBmixingEqn}) completely in terms of the invariant
overall rest energy $M$ and the masses $m_W$ and $m_B=i Q_B$.

\section{$W_0 + B_0$ Mixing into Electro-Weak Bosons}
\indent \indent
Physical $W$, $Z$, and $H$ bosons undergo processes that obey kinematic conservation conditions. 
The strategy will be to examine combinations of various parameters of mixing
associated with a co-moving $H$, since
co-moving spinors have identical components, independent of mass.

\subsection{$W+H$ elastic scattering}
\indent \indent
As previously mentioned, meaningful relationships amongst the mixing parameters
are expected to imply on-shell conservation
satisfying $p_{1i}^\mu+p_{2i}^\mu=p_{1f}^\mu+p_{2f}^\mu$. 
Consider the scattering $W + H \rightarrow W  + H$.  
In the center-of-momentum frame, $p_H=-p_W$.
This can be expressed using (\ref{zetamEqn})
as a relationship between $\zeta_H$ and $\zeta_W$ given by
\be
\zeta_H (\zeta_W) = -\sin^{-1} \left (
m_W \sin \zeta_W \over \sqrt{m_H^2 \cos (2 \zeta_W) + 2 m_W^2 \sin^2 \zeta_W}
\right ).
\label{ZetaHofZetaWEqn}
\ee
Appropriate forms for $\zeta W$ and $\zeta_Z$ in (\ref{PhiWBeqn}) should be ascertained.

A particular $W_0-B_0$ orthogonal mixing, having overall invariant mass $M=m_Z$
with $m_B\rightarrow i Q_B = i m_H$, involves all electro-weak mass parameters,
and can be constructed from (\ref{zetaWBmixEqn}), resulting in the form
\be
\zeta W^{m_Z}_{m_H}=\sin^{-1} \left (
{\sqrt{-2 m_H m_W + \sqrt{(m_H^2 + m_W^2)^2 + 2 (m_H^2 - m_W^2) m_Z^2 + m_Z^4}  } \over
\sqrt{2} \left [ (m_H^2 + m_W^2)^2 + 2  (m_H^2 - m_W^2) m_Z^2 + m_Z^4
\right ]^{1 \over 4}  } \,
\right ).
\label{zetaWmZmHEqn}
\ee
Direct substitution of (\ref{zetaWmZmHEqn}) into (\ref{ZetaHofZetaWEqn})
results in the expression
\be
\begin{array}{l}
\zeta_H (\zeta W^{m_Z}_{m_H})= \\
\quad  -\sin^{-1} \left (
{1 \over \sqrt{2}} \sqrt{m_W \left (  -2 m_H m_W +
\sqrt{(m_H^2 + m_W^2)^2 + 2 (m_H^2 - m_W^2) m_Z ^2 + m_Z^4 }      \right )  \over
2 m_H^3  + m_W \left (  -2 m_H m_W +
\sqrt{(m_H^2 + m_W^2)^2 + 2 (m_H^2 - m_W^2) m_Z ^2 + m_Z^4 }      \right )      } \,
\right )
\end{array}.
\label{zetaHofzetaWmZmHEqn}
\ee
For this particular value of $\zeta_H$, the momenta
indeed satisfy $p_H (\zeta_H)=-p_W (\zeta W^{m_Z}_{m_H})$,
providing kinematically consistent values of opposing momenta
with energy conservation for elastic scattering of $H+H$, $W+W$, and $W+H$.

Finally, the mixing angle in (\ref{PhiWBeqn}) must be determined.
The values for  $\zeta W$ and $\zeta_Z$ will be chosen so that the mixing
parameters are co-moving with the $H$'s according to
$\zeta W= -\zeta_H (\zeta W^{m_Z}_{m_H})=-\zeta_Z$,
as demonstrated in Figure \ref{WZHtoWZH}.
\onefigure{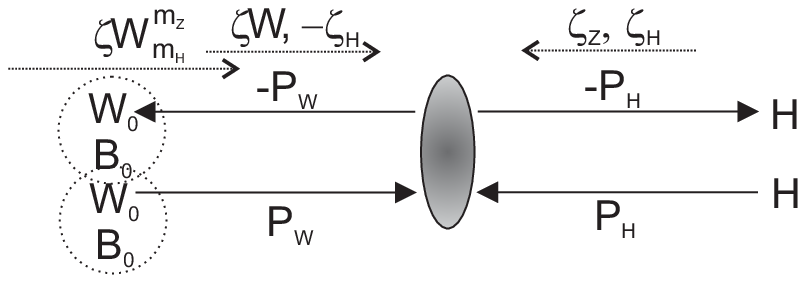}{$W_0-B_0$ mixings $\zeta W^{m_Z}_{m_H}$ (into invariant
mass $m_Z$) that are kinematically
consistent for $H$ elastic scattering with the $W$, satisfying $|P_W|=|P_H|$.  Momenta are
represented using solid arrows. The $H$'s simultaneously co-move with the mixing according to
$\zeta W$=$-\zeta_H=-\zeta_Z$ , as represented by the dashed arrows.}

The resulting mixing angle satisfies
\bea
\cos \phi_{WB}^{(WH)}  \Rightarrow  {2 m_H^3 \over
2 m_H^3 -2 m_H  m_W^2 +m_W \sqrt{(m_H^2 +m_W^2)^2 + 2 (m_H^2 -m_W^2) m_Z^2 + m_Z^4} } \nonumber  \\
= {2 \mu_{HZ}^3 \over 2  \mu_{HZ}^3 - 2 \mu_{HZ} \, \mu_{WZ}^2  +
\mu_{WZ} \sqrt{\mu_{HZ}^4 +(1 - \mu_{WZ}^2)^2 + 2 \mu_{HZ}^2 (1+\mu_{WZ}^2)  }  },  \qquad  \label{CosWZHWZHEqn}
\eea
where the mass ratios are defined by $\mu_{WZ}\equiv {m_W \over m_Z}$ and
$\mu_{HZ}\equiv {m_H \over m_Z}$.
Substitution of recent values of the electro-weak boson masses
from the \emph{Particle Data Group}  (PDG)\cite{PDG} gives a mixing angle
satisfying $\cos \phi_{WB} \simeq 0.8814644$, with deviation from the quoted values of 
${\Delta \cos \phi_{WB} \over \mu_{WZ} }\simeq -4.7 \times 10^{-6}$,
well within the experimental uncertainty.

\subsection{$W^+ W^-$, $Z Z$, $H H$ scattering re-arrangements}
\indent \indent
Next, consider scatterings involving boson pairs
$Z +\bar{Z}$, $W +\bar{W}$, and $H + \bar{H}$.  
For a given $H$ momentum $p_H$, the kinematics constrains $p_Z$, defining
$\zeta_Z (\zeta_H)$ given by
\be
\zeta_Z (\zeta_H) = \sin^{-1} \left (
{1 \over \sqrt{2}} \sqrt{m_H^2 +(m_H^2 -2 m_Z^2) \cos(2 \zeta_H) \over
m_H^2 + (m_H^2 - m_Z^2) \cos(2 \zeta_H)} \,
\right).
\label{zetaZofzetaHeqn}
\ee
Thus, kinematic systems co-moving with the $H$'s will
be used to specify the parameter $\zeta_Z$ in (\ref{PhiWBeqn}).

As was done in the previous section, the kinematic system should be
chosen to mix parameters involving $W$, $Z$, and $H$ in a meaningful manner
such that energy-momentum is conserved.  Consider
the momentum $p_{W^*}$ for $W +\bar{W}$ scattering in the rest frame of the $H$'s
(with energies $\epsilon_{W^*}=m_H$).  This
defines $\zeta_{W^*}$ using an equation analogous to
(\ref{zetaZofzetaHeqn}) given by
\be
\zeta_{W^*}=\sin^{-1} \left(  
{1 \over \sqrt{2}} \sqrt{2 m_H^2 - 2 m_W^2 \over 2 m_H^2 - m_W^2}
\right ),
\label{zetaWstarEqn}
\ee
as demonstrated in Figure \ref{WbarW}.
\onefigure{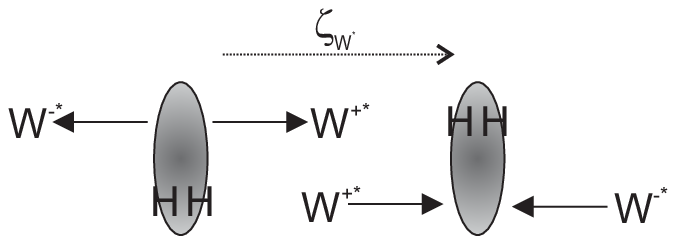}{Kinematic diagram for $W \bar{W}$ creation/annihilation in $H$'s rest frame.}
Next, consider a system involving $Z + \bar{Z}$ and $H + \bar{H}$.  If the $H$ in \emph{this} system
co-moves with an aforementioned $W^*$, i.e. $\zeta_H=\zeta_{W^*}$, then substituting 
(\ref{zetaWstarEqn}) into (\ref{zetaZofzetaHeqn}) results in the expression
\be
\zeta_Z (\zeta_H (\zeta_{W^*})) = \sin^{-1} 
\sqrt{1 - {m_H^4 \over  2 m_H^4 - m_W^2 m_Z^2}}.
\label{zetaZofzetaHofzetaWstarEqn}
\ee
This relationship indeed satisfies the kinematic requirement
$\epsilon_Z=\epsilon_H \rightarrow {m_H^2 \over m_W}$ in the center-of-momentum system. 
These parameters are demonstrated in Figure \ref{ZtoH}.
\onefigure{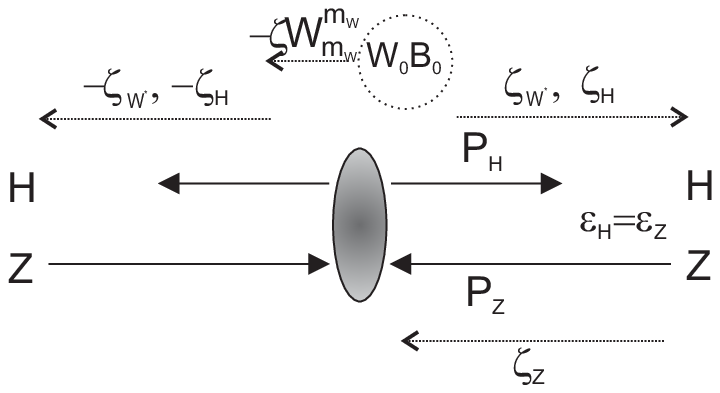}{Kinematic diagrams involving $Z,\bar{Z}, H$, and $\bar{H}$. The on-shell
$H$ is co-moving with a kinematically-consistent $W$ ($\zeta_H=\zeta_{W^*}$),
which coincides with energy-momentum
conservation $\epsilon_H=\epsilon_Z$ in the center-of-momentum system.}

Finally, an appropriate parameter $\zeta W ^M _{Q_B}$ for the $W_0+B_0$ will be identified
with $\zeta W$ in (\ref{PhiWBeqn}).  Since all of the kinematically-consistent mixing of
the bosons have been incorporated in $\zeta_Z$, the mixing in (\ref{zetaWBmixEqn})
should involve orthogonal $W$ mixing into $W$, with
$M=m_W, Q_B=m_W$. This gives the only (finite mass) mixing that results in a
mass-independent pure number:
\be
\zeta W ^{m_W} _{m_W}=\sin^{-1} \left (
\sqrt{{1 \over 2}- {1 \over \sqrt{5}}} \,
\right ).
\label{zetaWmWmWeqn}
\ee
Furthermore, this is the \emph{only} mixing form $\zeta W^M_{Q_B}$
(constructed from available mass parameters) indirectly
generated using the forms (\ref{zetamEqn})
for $\zeta_W$ and $\zeta_Z$ (substituted into (\ref{PhiWBeqn})) 
with $\epsilon_Z=\epsilon_H = {m_H^2 \over m_W}$ (and
consistent momenta), that directly results in a value for
$\cos \phi_{WB}$ consistent with ${m_W \over m_Z}$. 
Using (\ref{zetaZofzetaHofzetaWstarEqn}) and (\ref{zetaWmWmWeqn}) to calculate the mixing
angle in (\ref{PhiWBeqn}) gives the result
\bea
\cos \phi_{WB}   \goesto{Z\bar{Z}}{H\bar{H}} { \sqrt{50+20 \sqrt{5}} \: m_H^2 +
\sqrt{50-20 \sqrt{5}}\sqrt{m_H^4 - m_W^2 m_Z^2} \over
10 \sqrt{2 m_H^4  - m_W^2 m_Z^2} } \nonumber \\
= {\sqrt{5 + 2 \sqrt{5}} \, \mu_{HZ}^2 +  \sqrt{5 - 2 \sqrt{5}} \sqrt{\mu_{HZ}^4 -\mu_{WZ}^2} \over
\sqrt{20 \mu_{HZ}^4 -10 \mu_{WZ}^2} }.
\label{CosWBWWtoZZeqn}
\eea
In this case, substitution of mass values from the
PDG\cite{PDG} results in a mixing angle
of $\cos \phi_{WB} \simeq 0.8814718$, with deviation from the quoted values of 
${\Delta \cos \phi_{WB} \over \mu_{WZ} }\simeq 3.7 \times 10^{-6}$, also
well within the experimental uncertainty.

\subsection{Calculation of the mixing angle and mass ratios}
\indent \indent
As can be seen from the form of the equations (\ref{CosWZHWZHEqn}) and (\ref{CosWBWWtoZZeqn}),
the mixing angle and mass ratios are completely independent of the overall mass scale.
The H-Z mass ratio can be expressed in the closed form
\bea
\mu_{HZ} ={1 \over \sqrt{2}} \left (
{ \mu_{WZ}^2 \left [2 - \sqrt{5} + 2 (-5 + \sqrt{5}) \mu_{WZ}^2 + 10 \mu_{WZ}^4 +
\mu_{WZ}\sqrt{5-5 \mu_{WZ}^2}  \right ] \over
1 - 5 \mu_{WZ}^2 + 5 \mu_{WZ}^4  }
\right )^{1 \over 4}  \nonumber \\
\goesto{\cos \phi_{WB}}{{m_W \over m_Z} } \sqrt{\cos \phi_{WB}}  \left [ {1 \over 2} \left (
1+{ \sqrt{5} \over 2 \cos (2 \phi_{WB}) - \sin  (2 \phi_{WB})}
\right ) \right ]^{1 \over 4} . \qquad \qquad  \qquad   \label{HiggsMassEqn}
\eea

Although a closed form expression for $\mu_{WZ}$ has yet to be obtained, this pure number
is a direct prediction from consistency of (\ref{CosWZHWZHEqn}) and (\ref{CosWBWWtoZZeqn}). 
Choosing $m_Z$ as the most precisely measured mass, the predicted mass values are given by
\be
\begin{array}{l}
{m_W \over m_Z}= 0.8814666395389308...\\
{m_H \over m_Z}=1.3719494526608669... \\
\phi_{WB}=0.4918374045660035... \\
m_Z= 91.1876 \pm 0.0021 GeV\\
m_W=80.3788 \pm 0.0019 GeV\\
m_H=125.1048 \pm 0.0029 GeV.
\end{array}
\label{JLvaluesEqn}
\ee
These values fall well within the present uncertainties in
measurements of the mass ratios reported by the Particle Data Group\cite{PDG}:
\be
\begin{array}{l}
{m_W \over m_Z}= 0.88147 \pm 0.00013, \\
\theta_{WZ}=0.491830 \mp 0.000275, \\
m_Z= 91.1876 \pm 0.0021 GeV, \\
m_W=80.379 \pm 0.012 GeV, \\
m_H=125.10 \pm 0.14 GeV.
\end{array}
\label{PDGvaluesEqn}
\ee
Since only on-shell mass ratios are predicted, if the formulation
is indeed physically meaningful, any
deviations can only come from incorrect associations in the kinematic
processes presented.  Other meaningful processes must necessarily be
redundant.


\section{Discussion and Conclusions}
\indent \indent
Causal spinor fields have fundamental $\Gamma={1 \over 2}$
fermion representations with the correct number and types of gauge-field
generated interactions, while requiring any geometric interaction that is covariant
under curvilinear coordinate transformations must incorporate those interactions via the
principle of equivalence.
Unitary mixing of degenerate $\Gamma=1, \gamma=0$ boson spinors consistent with
on-shell kinematics has been utilized to determine
the mixing angle in terms of mass ratios. 
This allows all boson masses to be determined by a single particle mass scale.

The formulation has been applied using the experimental masses of the electro-weak
bosons.  By requiring a co-moving $H$ spinor associated with each mixing,
its vector ($J=1$) components can be associated with the identical
spinor components of the associated \emph{massive} gauge boson.  This
leaves only the scalar ($J=0$) component to associate with the Higgs boson.
The predicted mass ratios have been shown to be consistent with measured values.

The formulation also suggests the possibility of a $\mathbf{\Psi}^{(1)}_0 (J=0)$ electroweak
boson as a dark matter candidate.
Promising expressions for the "effective" mixing angle and its energy-scale dependence,
as well as additional relations amongst particle couplings,
are actively being explored as on-going research.

\subsection*{Acknowledgement}

The author is grateful for the unwavering support of Eileen Johnston throughout much of
this extensive effort.


\renewcommand{\theequation}{A.\arabic{equation}}
\setcounter{equation}{0}

\pagestyle{empty}
\section*{Appendix}
\subsubsection*{Matrix Representation of $\Gamma=1$ Systems}
\indent

The $\Gamma=1$ representations of the various group generators consist
of $10 \times 10$ matrices.  These matrices satisfy the Lorentz algebra
extended to include the relationships 
$
\left [ \Gamma^0 \, , \, \Gamma^k \right] \: = \: i \, K_k  , \: \: \:
\left [ \Gamma^0 \, , \, J_k \right] \: = \: 0  , \: \: \:
\left [ \Gamma^0 \, , \, K_k \right] \: = \: -i \,  \Gamma^k  , \: \: \:
\left [ \Gamma^j \, , \, \Gamma^k \right] \: = \: -i \, \epsilon_{j k m} \, J_m  , \\
\left [ \Gamma^j \, , \, J_k \right] \: = \: i \, \epsilon_{j k m} \, \Gamma^m  , \: \: \:
\left [ \Gamma^j \, , \, K_k \right] \: = \: -i \, \delta_{j k} \, \Gamma^0  .
$
Choosing the order of quantum numbers in the columns of the matrices as
$\left ( \begin{array}{c }  \gamma \\ J \\ s_z   \end{array} \right ) \rightarrow
\left \{
\left ( \begin{array}{r}  0 \\ 0 \\ 0   \end{array} \right ) , \right . \\
\left ( \begin{array}{r}  1 \\ 1 \\ 1   \end{array} \right ) ,
\left ( \begin{array}{c }  1 \\ 1 \\ 0   \end{array} \right ) ,
\left ( \begin{array}{r}  1 \\ 1 \\ -1   \end{array} \right ) ,
\left ( \begin{array}{r}  0 \\ 1 \\ 1   \end{array} \right ) ,
\left .  \left ( \begin{array}{r}  0 \\ 1 \\ 0   \end{array} \right ) ,
\left ( \begin{array}{r}  0 \\ 1 \\ -1   \end{array} \right ) ,
\left ( \begin{array}{r}  -1 \\ 1 \\ 1   \end{array} \right ) ,
\left ( \begin{array}{r}  -1 \\ 1 \\ 0   \end{array} \right ) ,
\left ( \begin{array}{r}  -1 \\ 1 \\ -1   \end{array} \right ) 
\right \} $: \\
\be
\begin{array}{l}
\mathbf{\Gamma}^0 \,=\, \left ( \begin{array}{r r r r r r r r r r}
0 & 0 & 0 & 0 & 0 & 0 & 0 & 0 & 0 & 0 \\
0 & 1 & 0 & 0 & 0 & 0 & 0 & 0 & 0 & 0 \\
0 & 0 & 1 & 0 & 0 & 0 & 0 & 0 & 0 & 0 \\
0 & 0 & 0 & 1 & 0 & 0 & 0 & 0 & 0 & 0 \\
0 & 0 & 0 & 0 & 0 & 0 & 0 & 0 & 0 & 0 \\
0 & 0 & 0 & 0 & 0 & 0 & 0 & 0 & 0 & 0 \\
0 & 0 & 0 & 0 & 0 & 0 & 0 & 0 & 0 & 0 \\
0 & 0 & 0 & 0 & 0 & 0 & 0 & -1 & 0 & 0 \\
0 & 0 & 0 & 0 & 0 & 0 & 0 & 0 & -1 & 0 \\
0 & 0 & 0 & 0 & 0 & 0 & 0 & 0 & 0 & -1 
\end{array} \right )  \\ \\
\mathbf{\Gamma}^x \,=\,   { 1 \over 2} \left ( \begin{array}{r r r r r r r r r r}
0 & -1 & 0 & 1 & 0 & 0 & 0 & -1 & 0 & 1 \\
1 & 0 & 0 & 0 & 0 & 1 & 0 & 0 & 0 & 0 \\
0 & 0 & 0 & 0 & 1 & 0 & 1 & 0 & 0 & 0 \\
-1 & 0 & 0 & 0 & 0 & 1 & 0 & 0 & 0 & 0 \\
0 & 0 & -1 & 0 & 0 & 0 & 0 & 0 & 1 & 0 \\
0 & -1 & 0 & -1 & 0 & 0 & 0 & 1 & 0 & 1 \\
0 & 0 & -1 & 0 & 0 & 0 & 0 & 0 & 1 & 0 \\
1 & 0 & 0 & 0 & 0 & -1 & 0 & 0 & 0 & 0 \\
0 & 0 & 0 & 0 & -1 & 0 & -1 & 0 & 0 & 0 \\
-1 & 0 & 0 & 0 & 0 & -1 & 0 & 0 & 0 & 0 
\end{array} \right )  \\ \\
\mathbf{\Gamma}^y \,=\,   { i \over 2} \left ( \begin{array}{r r r r r r r r r r}
0 & -1 & 0 & -1 & 0 & 0 & 0 & -1 & 0 & -1 \\
-1 & 0 & 0 & 0 & 0 & -1 & 0 & 0 & 0 & 0 \\
0 & 0 & 0 & 0 & 1 & 0 & -1 & 0 & 0 & 0 \\
-1 & 0 & 0 & 0 & 0 & 1 & 0 & 0 & 0 & 0 \\
0 & 0 & 1 & 0 & 0 & 0 & 0 & 0 & -1 & 0 \\
0 & -1 & 0 & 1 & 0 & 0 & 0 & 1 & 0 & -1 \\
0 & 0 & -1 & 0 & 0 & 0 & 0 & 0 & 1 & 0 \\
-1 & 0 & 0 & 0 & 0 & 1 & 0 & 0 & 0 & 0 \\
0 & 0 & 0 & 0 & -1 & 0 & 1 & 0 & 0 & 0 \\
-1 & 0 & 0 & 0 & 0 & -1 & 0 & 0 & 0 & 0 
\end{array} \right )  \\ \\
\mathbf{\Gamma}^z \,=\,   { 1 \over \sqrt{2} } \left ( \begin{array}{r r r r r r r r r r}
0 & 0 & 1 & 0 & 0 & 0 & 0 & 0 & 1 & 0 \\
0 & 0 & 0 & 0 &1 & 0 & 0 & 0 & 0 & 0 \\
-1 & 0 & 0 & 0 & 0 & 0 & 0 & 0 & 0 & 0 \\
0 & 0 & 0 & 0 & 0 & 0 & -1 & 0 & 0 & 0 \\
0 & -1 & 0 & 0 & 0 & 0 & 0 & 1 & 0 & 0 \\
0 & 0 & 0 & 0 & 0 & 0 & 0 & 0 & 0 & 0 \\
0 & 0 & 0 & 1 & 0 & 0 & 0 & 0 & 0 & -1 \\
0 & 0 & 0 & 0 & -1 & 0 & 0 & 0 & 0 & 0 \\
-1 & 0 & 0 & 0 & 0 & 0 & 0 & 0 & 0 & 0 \\
0 & 0 & 0 & 0 & 0 & 0 & 1 & 0 & 0 & 0 
\end{array} \right )  
\end{array}
\label{GammaMatrices}
\ee

The representation transforms a scalar ($J=0, \, \gamma=0$) component, as well as vector ($J=1$)
and ``iso-vector" ($\gamma=1,0,-1$) components.
For present purposes, the 10-spinors will be Hermitian normalized  (rather than ``Dirac" normalized) such that
for states labeled by $\gamma$ (the eigenvalue of $\mathbf{\Gamma}^0$),
\be
\mathbf{\Psi}_{\gamma'}^{(1)\dagger} ( \vec{p}, J', s_z ') ~ 
\mathbf{\Psi}_{\gamma}^{(1)} ( \vec{p}, J,  s_z) ~=~ 
\delta_{\gamma'  \gamma} \, \delta_{J'  J} \, \delta_{s_z '  s_z} .
\ee
The spinors satisfy the eigenvalue equation
\be
\mathbf{\Gamma}^\mu \hat{P}_\mu \mathbf{\Psi}_{\gamma}^{(1)} ( \vec{p},J, s_z) \: = \:
-\gamma  m c \,\mathbf{\Psi}_{\gamma}^{(1)} ( \vec{p},J, s_z) ,
\label{SpinorEigenvalueEqn}
\ee
with $m > 0$ \emph{always}.  For pairwise kinematics, it is sufficient to consider only z-moving particles,
since general states can be developed via rotation of the z-axis.
The normalized degenerate ($\gamma$=0) eigenstates take the form\\
$\{\mathbf{\Psi}_{0}^{(1)} ( \vec{p},0, 0),\mathbf{\Psi}_{0}^{(1)} ( \vec{p},1, 1),\mathbf{\Psi}_{0}^{(1)} ( \vec{p},1, 0),\mathbf{\Psi}_{0}^{(1)} ( \vec{p},1, -1)\}\Rightarrow$
\be
\left \{
\begin{array}{c c c c}
\left (
\begin{array}{c}
\sqrt{m^2+p_z^2} \over \sqrt{m^2+2 p_z^2}  \\ 0 \\ - p_z \over \sqrt{2} \sqrt{m^2+2 p_z^2}  \\
0 \\ 0 \\ 0 \\ 0 \\ 0 \\  p_z \over \sqrt{2} \sqrt{m^2+2 p_z^2}  \\ 0
\end{array}
\right ),  &
\left (
\begin{array}{c}
0 \\ p_z \over \sqrt{2} \sqrt{m^2+2 p_z^2}  \\ 0 \\ 0 \\  \sqrt{m^2+p_z^2} \over \sqrt{m^2+2 p_z^2} \\
0  \\ 0 \\ p_z \over \sqrt{2} \sqrt{m^2+2 p_z^2}  \\ 0 \\ 0 
\end{array}
\right ),  &
\left (
\begin{array}{c}
0 \\ 0  \\ 0 \\ 0 \\ 0 \\ 1 \\ 0 \\ 0 \\ 0 \\ 0
\end{array}
\right ),  &
\left (
\begin{array}{c}
0 \\ 0 \\ 0 \\ -p_z \over \sqrt{2} \sqrt{m^2+2 p_z^2}  \\ 0 \\ 0 \\  \sqrt{m^2+p_z^2} \over \sqrt{m^2+2 p_z^2} \\ 0  \\ 0 \\ -p_z \over \sqrt{2} \sqrt{m^2+2 p_z^2}
\end{array}
\right )
\end{array}
\right \}
\label{DegenerateSpinors}
\ee
Degenerate eigenstates labeled with differing $m$ all satisfy (\ref{SpinorEigenvalueEqn}) with its
right-hand side set to zero, and thus they can freely mix. 
For completeness, spinors $\{\mathbf{\Psi}_{1}^{(1)} ( \vec{p},1, 1),
\mathbf{\Psi}_{1}^{(1)} ( \vec{p},1, 0), \mathbf{\Psi}_{-1}^{(1)} ( \vec{p},1, 0)   \}$
are displayed:
\be
\left \{
\left (
\begin{array}{c}
0 \\ {1 \over 2} \left ( 1+{m \over \sqrt{m^2+p_z^2}} \right )  \\ 0 \\ 0 \\
p_z \over \sqrt{2} \sqrt{m^2+p_z^2}  \\ 0 \\ 0 \\ 
{1 \over 2} \left ( 1-{m \over \sqrt{m^2+p_z^2}} \right )   \\ 0 \\ 0
\end{array}
\right ), \:
\left (
\begin{array}{c}
 - p_z \over \sqrt{2} \sqrt{m^2+p_z^2} \\ 0 \\ {1 \over 2} \left ( 1+{m \over \sqrt{m^2+p_z^2}} \right )  \\
0 \\ 0 \\ 0 \\ 0 \\ 0 \\  -{1 \over 2} \left ( 1-{m \over \sqrt{m^2+p_z^2}} \right )   \\ 0
\end{array}
\right ), \:
\left (
\begin{array}{c}
  p_z \over \sqrt{2} \sqrt{m^2+p_z^2} \\ 0 \\ -{1 \over 2} \left ( 1-{m \over \sqrt{m^2+p_z^2}} \right )  \\
0 \\ 0 \\ 0 \\ 0 \\ 0 \\  {1 \over 2} \left ( 1+{m \over \sqrt{m^2+p_z^2}} \right )   \\ 0
\end{array}
\right )
\right \}.
\label{TypicalSpinors}
\ee


\end{document}